\documentclass[12pt]{JHEP3}
\usepackage{amsmath,amsfonts,amssymb}
\usepackage{verbatim,graphicx,float}  

\title{Gravitational waves from the Big Bounce}

\author{Jakub Mielczarek \\ 
Marc Kac Complex Systems Research Centre, Jagiellonian University,\\
Reymonta 4, 30-059 Cracow, Poland 
\\

\email{jakub.f.mielczarek@gmail.com} \\
}

\abstract{
In this paper we investigate gravitational waves production during the Big Bounce phase inspired
by the Loop Quantum Cosmology. We consider the influence of the holonomy corrections
to the equation for tensor modes. We show that they act like additional effective 
graviton mass, suppressing gravitational waves creation. However, this effects can be 
treated perturbatively. We investigate the simplified model without these corrections
and find its exact analytical solution. For this model we calculate a spectrum
of the gravitational waves from the Big Bounce phase. The obtained spectrum 
decreases to zero for the low energy modes. Based on this observation we 
indicate that this effect can lead to the low CMB multipoles suppression and 
gives a potential way to test Loop Quantum Cosmology models. 
We also consider a scenario with a post-bounce inflationary phase. The obtained power spectrum
gives qualitative explanation of the CMB spectra, including low multipoles suppression.
This result is a challenge to construct a consistent bounce+inflation model 
in the Loop Quantum Cosmology.
}

\begin{document}

\section{Introduction} \label{sec:Intro}

Gravitational waves seems to be the best tool to explore early universe.
In particular they can be potentially used to verify quantum cosmological models.
This idea is based on the fact that gravitational waves produced during the quantum epoch can survive 
frozen on the super-horizontal scales. Then after re-entering  the horizon they can give imprint
on the CMB spectrum, which is observed today. From the empirical point of view, 
the most promising is the B spectrum of the CMB polarization. This spectrum 
has its source only in the tensor part of perturbation and when observed 
gives the direct method to investigate relic gravitational waves.
It is expected that the PLANCK mission could give an opportunity 
to detect the B polarization \cite{:2006uk}. However, such predictions 
base on the simple inflationary models. 

One of the most promising approaches to quantize gravity is Loop Quantum Gravity \cite{Ashtekar:2004eh}.
Based on this the theory of the quantum universe, Loop Quantum Cosmology \cite{Bojowald:2006da} arose. 
This theory predicts that the initial singularity state is replaced by the quantum
Big Bounce \cite{Ashtekar:2006rx}. In this scenario a universe undergoes contraction and then after quantum bounce 
evolves toward an expanding phase. During the the bounce energy density reach maximal finite energy density $\rho_c$.
The phenomenological description of the bounce phase 
can be obtained from the modified Friedmann equation 
\begin{equation}
H^2 = \frac{\kappa}{3} \rho \left(1 -\frac{\rho}{\rho_{\text{c}}} \right). 
\label{Friedmann}
\end{equation}
where $\kappa=8\pi G$.

Investigation of the perturbations in the cosmological models are crucial from the point 
of a large scale structures creation and exploration of the early universe. 
In the Loop Quantum Cosmology this issue has been studied in Ref. 
\cite{Bojowald:2006tm,Bojowald:2006zb,Bojowald:2007ab,Bojowald:2008gz}.
In the present paper we consider a particular kind of metric perturbations, the gravitational waves.
The creation of gravitons in models inspired by the Loop Quantum Cosmology  
were initially studied in Ref. \cite{Mielczarek:2007zy,Mielczarek:2007wc}.
Then the equation for the tensor modes with holonomy and inverse volume 
corrections has been derived in Ref. \cite{Bojowald:2007cd}. Recently the equation with holonomy effects has been
applied to the inflationary phase \cite{Barrau:2008hi}. 

In the present paper we investigate the creation of the gravitational waves during the Big Bounce 
phase. In our considerations we take into account only holonomy effects. The inverse volume corrections
exhibit the fiducial cell dependence and are not adequate in the models with the flat background
\cite{Ashtekar:2006wn}. The production of perturbations in the bouncing cosmologies has 
been recently studied in the different context in Ref. 
\cite{Gasperini:2003pb,Creminelli:2007aq,Cai:2007zv,Novello:2008ra}.

\section{Tensor modes with holonomy corrections}

The equation for tensor modes with holonomy corrections has been derived by Bojowald and Hossain \cite{Bojowald:2007cd}.
In the source free case these equation takes a form  
\begin{equation}
\frac{d^2}{d\tau^2} h_{i} + 
2 \left( \frac{\sin 2\bar{\mu} \gamma \bar{k} }{2\bar{\mu}\gamma}\right)\frac{d}{d\tau}  h_{i}
 -\nabla^2 h_{i} +T_{Q}  h_{i} = 0 \label{modehol}
\end{equation}
where $i=\oplus,\otimes$ indicate polarization state and 
\begin{eqnarray}
T_{Q} =  -2 \left(  \frac{\bar{p}}{\bar{\mu}} \frac{\partial \bar{\mu}}{\partial \bar{p}} \right) \bar{\mu}^2 \gamma^2  
 \left[  \frac{\sin\left( \bar{\mu} \gamma \bar{k} \right)  }{\bar{\mu} \gamma } \right]^4.
\end{eqnarray}
Here $\bar{p}=a^2$ and $\bar{k}=\bar{p}'/2\bar{p}$. The Hamilton equation for the variable $\bar{p}$ takes a form 
\cite{Bojowald:2007cd}
\begin{equation}
{\bar{p}}' = 2 \bar{p}  
 \left( \frac{\sin 2 \bar{\mu} \gamma  \bar{k}}{2\bar{\mu} \gamma}\right),
\end{equation}
what leads to the 
\begin{equation}
\left( \frac{\sin 2 \bar{\mu} \gamma \bar{k} }{2 \bar{\mu} \gamma}\right) = \frac{a'}{a}.
\end{equation}
The above equality indicates that the friction term in equation (\ref{modehol}) holds 
classical form. The effects of holonomies is the additional correction term $T_Q$ to the classical 
equation for tensor modes. This factor acts like an additional effective graviton mass. 

To define the correction $T_Q$ we need to specify a function $\bar{\mu}$. 
In general there is some freedom of the choice of this function in the power law form.
However, it has been recently 
shown that for the flat FRW models the only consistent choice is \cite{Corichi:2008zb}  
\begin{equation}
\bar{\mu} = \sqrt{\frac{\Delta}{\bar{p}}}
\end{equation}
where $\Delta=2\sqrt{3}\pi \gamma l^2_{\text{Pl}}$, which is called a $\bar{\mu}$-scheme.

\begin{figure}[ht!]
\centering
\includegraphics[width=9cm,angle=0]{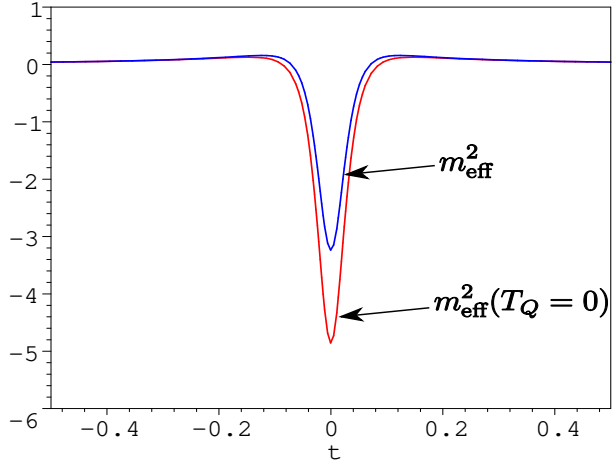}
\caption{Evolution of the effective masses $m^2_{\text{eff}}$  and $m^2_{\text{eff}}(T_Q=0)$.}
\label{meff}
\end{figure}

Introducing a new variable
\begin{equation}
u=\frac{a h_{\oplus}}{\sqrt{16\pi G}}=\frac{a h_{\otimes}}{\sqrt{16\pi G}} 
\end{equation}
and performing the Fourier transform
\begin{equation}
u(\tau,{\bf x}) = \int \frac{d^3{\bf k}}{(2\pi)^3}  u(\tau,{\bf k}) e^{i{\bf k}\cdot {\bf x}}
\end{equation}
we can rewrite the equation (\ref{modehol}) to the form
\begin{equation}
\frac{d^2}{d\tau^2} u(\tau,{\bf k}) +[k^2 +m^2_{\text{eff}}] u(\tau,{\bf k}) = 0 \label{ueq1}
\end{equation}
where $k^2={\bf k}\cdot {\bf k}$ and 
\begin{equation}
m^2_{\text{eff}}  = T_{Q} - \frac{a^{''}}{a}. \label{deffmeff}
\end{equation}
To calculate this function we must to specify the background dynamics.
We consider the model with a free scalar field. In this case evolution of the 
parameter $\bar{p}$ takes a form \cite{Mielczarek:2008zv}
\begin{equation}
\bar{p} = \left(A+Bt^2\right)^{1/3} \label{bouncesol}
\end{equation}
where
\begin{equation} 
A= \frac{1}{6} \kappa \pi^2_{\phi} \gamma^2 \Delta \  \ \text{and} \ \  B =\frac{3}{2}\kappa \pi^2_{\phi} . 
\end{equation}
Based on definition (\ref{deffmeff}) we calculate 

\begin{equation}
m^2_{\text{eff}} =  \frac{\kappa^2 \pi^4_{\phi}}{4}
\frac{\left(t^2-\frac{2}{9} \gamma^2\Delta \right)}{\left(A+Bt^2\right)^{5/3} }.  
\end{equation}
In the case $T_Q=0$ we obtain
\begin{equation}
m^2_{\text{eff}}(T_Q=0)= \frac{\kappa^2 \pi^4_{\phi}}{4}
\frac{\left(t^2-\frac{1}{3} \gamma^2\Delta \right)}{\left(A+Bt^2\right)^{5/3} }.  
\end{equation}
In Fig. \ref{meff} we show the evolution of the effective masses $m^2_{\text{eff}}$  and $m^2_{\text{eff}}(T_Q=0)$. 
Values of these functions are crucial from the point of view of gravitational waves production.
Generally, more negative this function is more gravitational waves is produced. 
We see that the presence of correction $T_Q$ leads to suppression of the gravitons production.
However, this effect is relatively weak.

For the bouncing universe considered we can write a closed system of equations
\begin{eqnarray}
\frac{du}{d\tau} &=& \pi_u, \\
\frac{d\pi_u}{d\tau} &=&- \left[ k^2+ \frac{\kappa^2 \pi^4_{\phi}}{4}
\frac{\left(t^2-\frac{2}{9} \gamma^2\Delta \right)}{\left(A+Bt^2\right)^{5/3} }  \right] u, \\ 
\frac{dt}{d\tau} &=& \left(A+Bt^2\right)^{1/6},
\end{eqnarray}
to describe the evolution of the gravitational waves. This system contains the holonomy correction 
to the background dynamics as well as to the correction to the equation for tensor modes.
Numerical solutions of this system can fully determinate the classical evolution of the gravitational
waves during the Big Bounce phase. 

\begin{figure}[ht!]
\centering
$\begin{array}{cc}   
\includegraphics[width=7cm,angle=0]{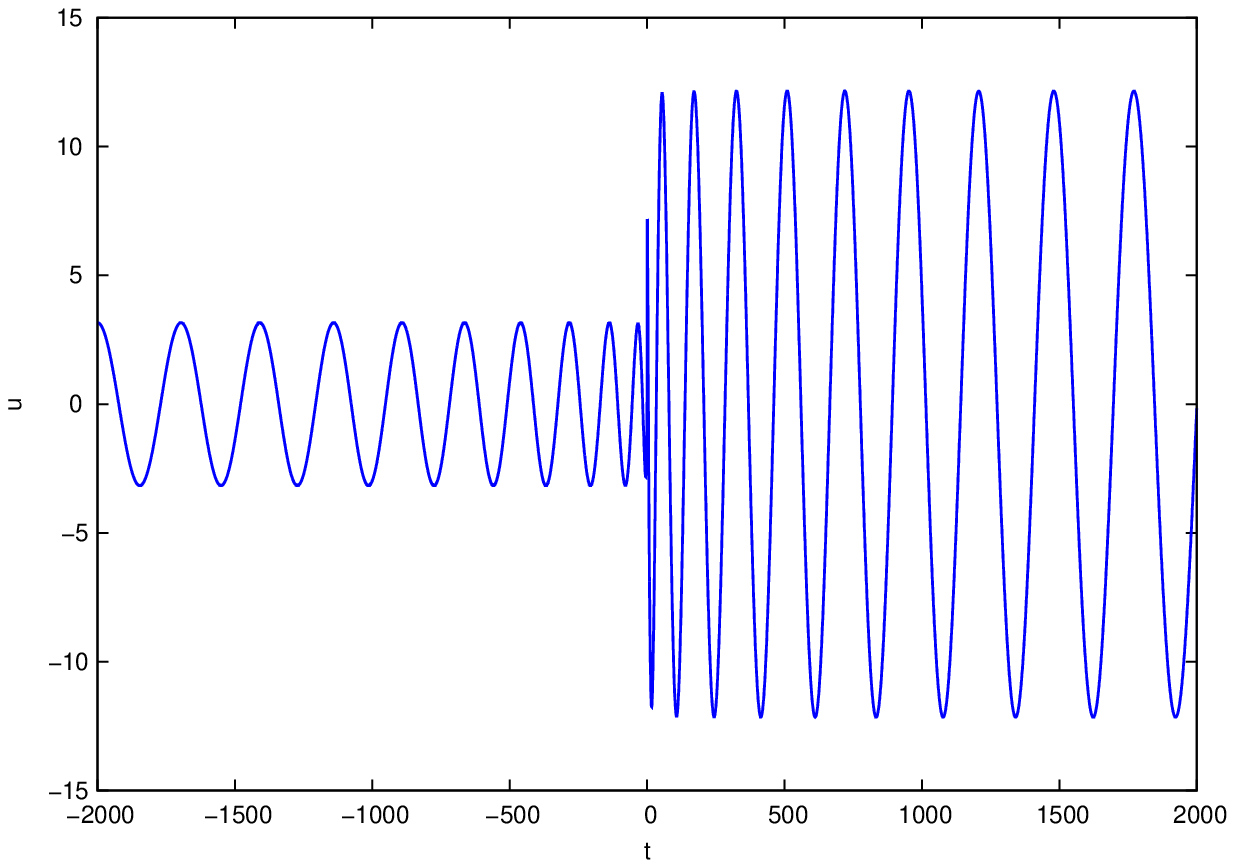}  &  \includegraphics[width=7.5cm,angle=0]{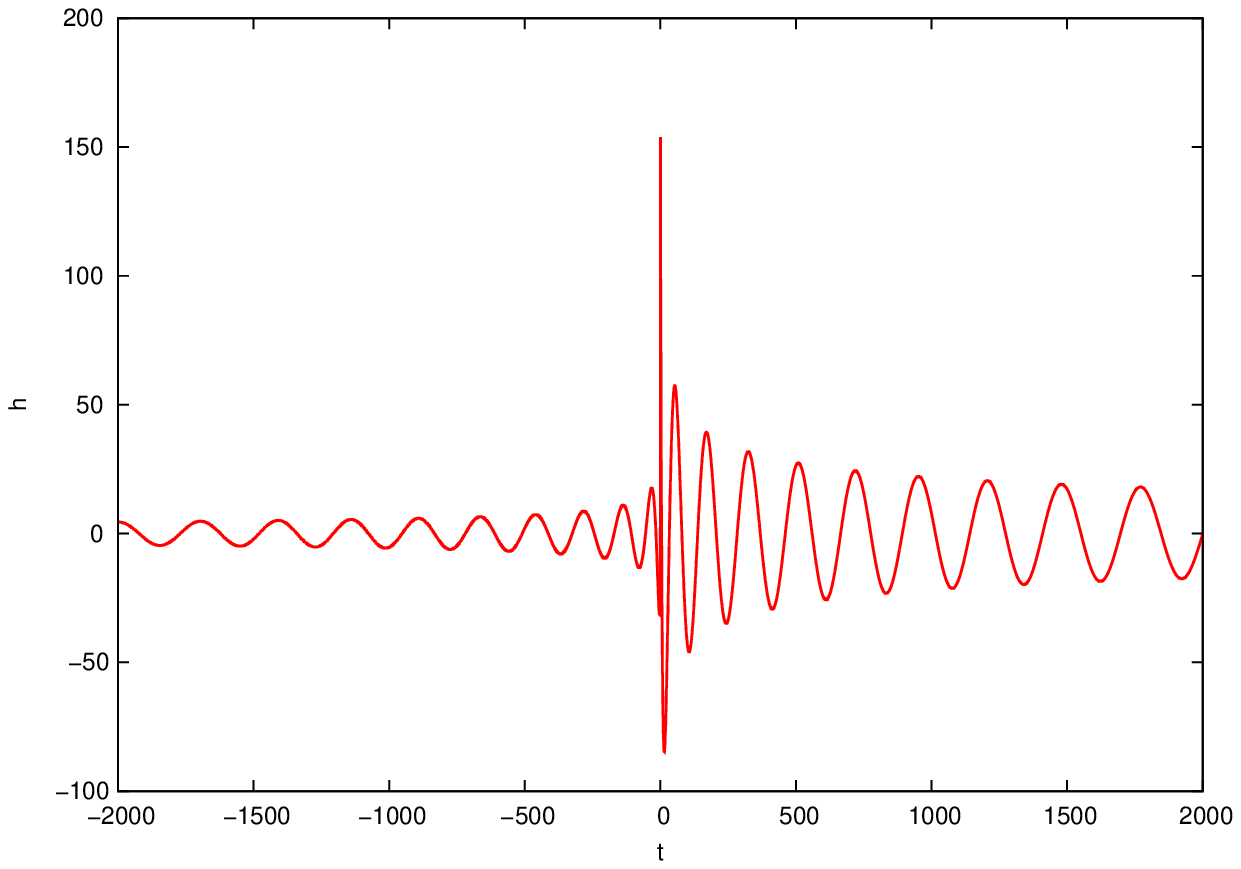}   
\end{array}
$
\caption{ 
{\bf Left  }: Evolution of the field $u$. 
{\bf Right }: Amplification of the tensor modes $h$ during the bounce.}
\label{EvolUHNUMERIC}
\end{figure}

In Fig. \ref{EvolUHNUMERIC} we show typical solutions for the function $u$ and  $h_i$ for 
the vacuum initial condition $u\sim 1/\sqrt{k}$. 
It is transparent that the tensor modes are amplified during the bounce phase.

In the next section we aim to determinate qualitative and quantitative
properties of the gravitons produced during the bounce phase.  

\section{Toy model}

The goal of this section is to give the introduction 
to the process of particles production during the bounce phase.
We show a simple, fully analytical model of the gravitational waves creation during the bounce. 
The model considered does not contain any quantum holonomy corrections to the mode equation.
This simplification can be however justified. Namely, as it was shown in the previous section,
the holonomy correction $T_Q$ leads only to the perturbative modifications.  

The model considered is based on the two assumptions: 
\begin{itemize}
\item we neglect quantum holonomy corrections $T_Q$ in the mode equation,
\item we assume  matter content in the form 
\begin{equation}
\rho = \frac{\rho_{\text{c}}}{a^2}.
\label{PFG:Matter}
\end{equation}
\end{itemize}

\subsection{Background dynamics}

Plugging condition (\ref{PFG:Matter}) to the modified Friedmann equation (\ref{Friedmann}) we find the 
solution 
\begin{equation}
a(t) =\sqrt{1+(t/t_0)^2}
\end{equation}   
where 
\begin{equation}
t_0^2 = \frac{3}{8\pi G \rho_{\text{c}}}. 
\end{equation} 
For the further calculations it will be useful to express 
this solution in terms of the conformal time. Performing   
transformation $d\tau= dt/a$ we find   
\begin{equation}
a(\tau) = \cosh (\tau/t_0).
\end{equation} 

\subsection{Mode functions}
Equation (\ref{ueq1}) without the term $T_Q$ takes a form
\begin{equation}
\frac{d^2}{d\tau^2} u(\tau,{\bf k}) +[k^2-\frac{a^{''}}{a}] u(\tau,{\bf k}) = 0. \label{umodeeq}
\end{equation}
This equation can be obtained from the action  
\begin{equation}
S_{\text{t}} = \frac{1}{2} \int d \tau d^3 {\bf x} [ u^{'2}-\delta^{ij}
 \partial_i u \partial_j u  -m^2_{\text{eff}}u^2 ] 
\end{equation}
where
\begin{equation}
m^2_{\text{eff}} = - \frac{a^{''}}{a}. 
\end{equation}
Canonical momenta conjugated to the variable $u$ we obtain from  
\begin{equation}
\pi(\tau,{\bf x})=\frac{\delta S_{\text{t}}}{\delta u'} = u'. 
\end{equation}

The next step is to quantise this theory. We perform canonical quantisation 
$(u,\pi) \rightarrow (\hat{u},\hat{\pi})$ introducing relations of the commutations
$[\hat{u}({\bf x},\tau),\hat{\pi}({\bf y},\tau)] = i \delta^{(3)} ({\bf x}-{\bf y})$  and 
$[\hat{u}({\bf x},\tau),\hat{u}({\bf y},\tau)] =[\hat{\pi}({\bf x},\tau),\hat{\pi}({\bf y},\tau)] =0$.
The operators $\hat{u},\hat{\pi}$ can be decomposed for the Fourier modes
\begin{eqnarray}
\hat{u}(\tau,{\bf x} )   &=& \frac{1}{2(2\pi)^{3/2}} \int d^3{\bf k } 
\left[ \hat{u}_{{\bf k}}(\tau) e^{i{\bf k}\cdot {\bf x}} +  
\hat{u}_{{\bf k}}^{\dagger}(\tau) e^{-i{\bf k}\cdot {\bf x}}  \right],   \label{TMdecomp1} \nonumber       \\
\hat{\pi}(\tau,{\bf x} ) &=& \frac{1}{2(2\pi)^{3/2}} \int d^3{\bf k }
 \left[ \hat{\pi}_{{\bf k}}(\tau) e^{i{\bf k}\cdot {\bf x}} +  
\hat{\pi}_{{\bf k}}^{\dagger}(\tau) e^{-i{\bf k}\cdot {\bf x}}  \right].  \label{TMdecomp2} \nonumber        
\end{eqnarray}
where 
\begin{eqnarray}
\hat{u}_{ {\bf k}  }(\tau)  &=&  \hat{a}_{{\bf k} } f(k,\tau)+ 
\hat{a}_{-{\bf k} }^{\dagger} f^{*}(k,\tau), \label{TMsol11}      \\
\hat{\pi}_{{\bf k} }(\tau)  &=&  \hat{a}_{{\bf k} } g(k,\tau)+ 
\hat{a}_{-{\bf k} }^{\dagger} g^{*}(k,\tau), \label{TMsol22}    
\end{eqnarray}
and $f(k,\tau)'=g(k,\tau)$.

The equation for the mode function takes a form
\begin{equation}
\frac{d^2}{d\tau^2}f(k,\tau) + \left[ k^2 +m^2_{\text{eff}} \right] f(k,\tau) = 0. \label{TMmodeeq}     
\end{equation}
where
\begin{equation}
m^2_{\text{eff}} = - \frac{a^{''}}{a} = -\frac{1}{t^2_0} =-k_0^2.
\end{equation}

With the use of definition
\begin{equation}
t_0^2 = \frac{3}{8\pi G \rho_{\text{c}}} \ \ 
 \text{and} \ \ \rho_{\text{c}} = \frac{\sqrt{3}}{16\pi^2\gamma^2l^4_{\text{Pl}}}.   
\end{equation}
and assuming  $\gamma=\gamma_M =0.12738$ \cite{Meissner:2004ju} we obtain 
\begin{equation}
k_0 \simeq \frac{2.38}{l_{\text{Pl}}}. 
\end{equation}

Now we can find solutions of equation (\ref{TMmodeeq}). We consider two cases 
$k^2>k^2_0$ and $k^2<k^2_0$.  
\begin{itemize}
\item Case $k^2>k_0^2$

The solution of equation (\ref{TMmodeeq}) takes a form
\begin{equation}
f(k,\tau) = Ae^{-i\Omega \tau}+ Be^{i\Omega \tau}
\end{equation}
where $A,B \in \mathbb{C}$ and
\begin{equation}
\Omega = \sqrt{k^2-k^2_0}.
\end{equation}
Performing the normalisation, with the use of the Wronskian condition, we obtain 
\begin{equation}
|A|^2-|B|^2=\frac{1}{2\Omega}. 
\end{equation}  
We choose advanced modes taking
\begin{equation} 
A= \frac{1}{\sqrt{2\Omega}} \  \ \text{and} \ \  B = 0,
\end{equation} 
what gives
\begin{eqnarray}
 f(k,\tau) &=&  \frac{1}{\sqrt{2\Omega}} e^{-i\Omega \tau},  \label{mode1}     \\
 g(k,\tau) &=& f^{'}(k,\tau) = -i\sqrt{\frac{\Omega}{2}} e^{-i\Omega \tau}.    
\end{eqnarray}

\item Case $k^2<k^2_0$

The solution of equation (\ref{TMmodeeq}) takes a form
\begin{equation}
f(k,\tau) = Ae^{-\bar{\Omega} \tau}+ Be^{\bar{\Omega} \tau}
\end{equation}
where $A,B \in \mathbb{C}$ and
\begin{equation}
\bar{\Omega} = \sqrt{k^2_0-k^2}.
\end{equation} 
Performing the normalisation, with the use of the Wronskian condition, we obtain 
\begin{equation}
BA^*-AB^* = -\frac{i}{2\bar{\Omega}} 
\end{equation}
what is fulfilled by
\begin{equation} 
A= \frac{i}{2\sqrt{\bar{\Omega}}} \  \ \text{and} \ \  B = \frac{1}{2\sqrt{\bar{\Omega}}}. 
\end{equation}
In this case we obtain
\begin{eqnarray}
 f(k,\tau) &=&  \frac{1}{2\sqrt{\bar{\Omega}}}  \left[ e^{\bar{\Omega}\tau} + i e^{-\bar{\Omega}\tau} \right], 
\label{mode2}      \\
 g(k,\tau) &=& f^{'}(k,\tau) = \frac{\sqrt{\bar{\Omega}}}{2}  \left[ e^{\bar{\Omega}\tau} - i e^{-\bar{\Omega}\tau} \right].   
\end{eqnarray}

\end{itemize}

\subsection{Power spectrum}

\begin{figure}[ht!]
\centering
\includegraphics[width=8cm,angle=0]{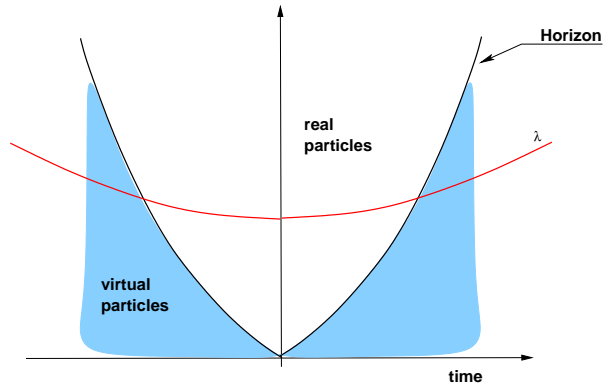}
\caption{Horizon in the bouncing cosmologies.}
\label{BounceHor}
\end{figure}

The correlation function for the tensor modes takes a form 
\begin{eqnarray}
\langle 0|  \hat{h}^a_b(\vec{x},\tau) \hat{h}^b_a (\vec{y},\tau)| 0 \rangle &=&
4 \frac{16\pi G}{a^2} \int \frac{d^3k}{(2\pi)^3} |f(k,\tau)|^2 e^{-i\vec{k}\cdot\vec{r}}  \nonumber  \\ 
 &=& \int \frac{dk}{k} \mathcal{P}_T(k,\tau) \frac{\sin kr}{kr},  
\end{eqnarray}
where we have defined the power spectrum
\begin{equation}
\mathcal{P}_T(k,\tau) = \frac{64\pi G}{a^2} \frac{k^3}{2\pi^2}  |f(k,\tau)|^2. \label{PowerSpectDef}
\end{equation}
Our goal now is to determinate this spectrum on the Hubble (horizon) scales. The evolution of the 
Hubble radius during the bouncing phase considered takes a form 
\begin{equation}
ds^2= 0 \ \ \rightarrow \ \ R_{\text{H}} = \pm a(\tau) \int_0^{\tau} d \tau'  =  \pm a(\tau) \tau 
\end{equation}
where  $ + $ denotes $expansion$ and $ - $ denotes $contraction$. We can also express it in term 
of the coordinate time $t$, obtaining
\begin{equation}
 R_{\text{H}} = \pm a(t) \int_0^{t}\frac{dt'}{a(t')} =   \pm t_0 \sqrt{1+(t/t_0)^2} \text{arcsh} (t/t_0).  
\end{equation} 
This function decreases to zero in the pre-Big Bang phase and increases in the post-Big Bang phase.
In Fig. \ref{BounceHor} we show evolution of the horizon for the typical bouncing cosmologies.
We draw also evolution of the arbitrary length scale $\lambda \propto a$.
We see that during the bouncing phase all length scales finally cross the horizon.
Quantum fluctuations of the length $\lambda < R_{\text{H}}$ behave like a virtual particles,
when fluctuations of the length $\lambda > R_{\text{H}}$ become classical excitations.

\begin{figure}[ht!]
\centering
\includegraphics[width=9cm,angle=0]{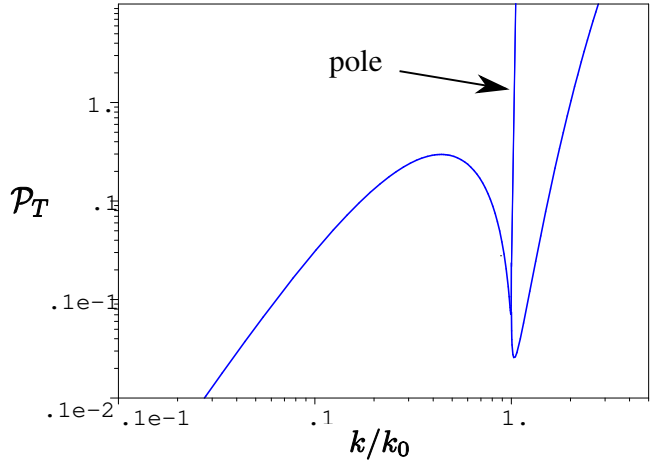}
\caption{Spectrum of the tensor perturbations on the horizon.}
\label{TensSpect}
\end{figure}

The condition $\lambda = R_{\text{H}}$ indicate 
\begin{equation} 
k= \frac{2\pi a}{R_{\text{H}}} =  \frac{2\pi a}{\pm \tau a} =  \frac{2\pi}{\pm \tau}. 
\end{equation}
Plugging it to definition (\ref{PowerSpectDef}) and using the mode functions (\ref{mode1}), (\ref{mode2})
we determinate the tensor power spectrum on the horizon scales
\begin{equation}
\mathcal{P}_T(k) = 
\left\{ \begin{array}{ccl} 
 \frac{16}{\pi} \left(\frac{k_0}{m_{\text{Pl}}}\right)^2 \left(\frac{k}{k_0}\right)^3 
 \frac{1}
{\sqrt{(k/k_0)^2-1} \text{ch}^2\left( \frac{2\pi}{k/k_0} \right) } & \mbox{for}  & k>k_0  \\
 \frac{16}{\pi} \left(\frac{k_0}{m_{\text{Pl}}}\right)^2 \left(\frac{k}{k_0}\right)^3 
 \frac{\text{ch} \left[ 4\pi  \sqrt{\left({k_0}/{k}\right)^2 -1  }   \right] }
{\sqrt{1-(k/k_0)^2} \text{ch}^2\left( \frac{2\pi}{k/k_0} \right) }    & \mbox{for}  & k < k_0
\end{array} \right. 
\end{equation}
We plot this spectrum in Fig. \ref{TensSpect}. 
We see that for $k<k_0$ the spectrum is characterised by the bump. For $k \rightarrow 0 $ 
the spectrum is decreasing to zero. However, for $k \rightarrow k_0 $ the spectrum 
has a pole $\mathcal{P}_T\rightarrow \infty$. This pole is however not physical. 
To show it we consider the correlation function
\begin{eqnarray}
\langle 0|  \hat{h}^a_b(\vec{x},\tau) \hat{h}^b_a (\vec{x},\tau)| 0 \rangle &=&
\frac{32 G}{\pi a^2} \int_0^{\infty} dk  k^2 |f(k,\tau)|^2,     
\end{eqnarray}
where 
\begin{eqnarray}
|f(k,\tau)|^2 &=& \Theta(k-k_0) \frac{1}{2\sqrt{k^2-k^2_0}} \nonumber  \\
 &+& \Theta(k_0-k) \frac{1}{2\sqrt{k^2_0-k^2}} 
\text{ch} \left[ 2\sqrt{k^2_0-k^2}\tau \right]. \nonumber
\end{eqnarray}
The correlation function is a physical quantity and can indicate which features 
of the spectrum $\mathcal{P}_T$ are real divergences. As we can see, near 
the pole $k=k_*$ the correlation function is determined by the integrals 
\begin{eqnarray}
I_1 = \int dk  \frac{k^{2}}{\sqrt{k^2-k^2_0}} = \left\{ x=\frac{k}{k_0}   \right\}  
    = \frac{k_0^2}{2} \left[ x\sqrt{x^2-1} - \ln \left|x+\sqrt{x^2-1} \right|  \right]  
\end{eqnarray}
and
\begin{eqnarray}
I_2 = \int dk  \frac{k^{2}}{\sqrt{k^2_0-k^2}} = \left\{ x=\frac{k}{k_0}   \right\} 
    = \frac{k_0^2}{2} \left[ -x\sqrt{1-x^2} - \text{arcsin} x  \right].     
\end{eqnarray}
Both of them are finite for the $x=1$ and give a finite contribution to the definition 
of the correlation function.

For $k \rightarrow \infty $ the power spectrum exhibit a UV divergence. In fact, the region $k>k_0$ traces the 
Planck scales and a different approach should be used to determinate properties 
of the quantum fluctuations. Another important issue is to indicate which of the 
modes survive frozen above the Hubble radius. In fact, not all of them but only 
these with $k<k_0$. It can be seen from the asymptotic solutions of 
equation (\ref{umodeeq}). Namely, in the regime  $k^2 \ll |a^{''}/a|=k_0^2$ we have solution
\begin{equation}
 h_i \simeq  A_k +B_k\int^{\tau} \frac{dx}{a^2(x)}. 
\end{equation}
Here we have a constant contribution $A_k$ which freezes the amplitude of the gravitational
waves. In this regime the spectrum does not change during the evolution above the horizon 
and will be the same when re-enters the horizon. 
In the regime $k^2 \gg |a^{''}/a|=k_0^2$  we have decaying solutions
\begin{equation}
h_i \simeq \frac{e^{\pm ik\tau}}{a}.
\end{equation}
These modes does not lead to the classical fluctuations when re-enter horizon.

Summing up, the spectrum $\mathcal{P}_T$ on the second branch of the horizon is 
free form the UV divergent part for the $k>k_0$. The only contribution to this spectrum 
comes from the bump for the $k<k_0$.  

\subsection{Parameter $\Omega_{\text{gw}}$}  

To describe the spectrum of gravitons it 
is common to use the parameter
\begin{equation}
\Omega_{\text{gw}}(\nu) =\frac{\nu}{\rho_*}\frac{d \rho_{\text{gw}}}{d \nu}
\label{omegaGW}
\end{equation}
where $\rho_{\text{gw}} $ is the energy density of gravitational waves 
and $\rho_*$ is present critical energy density. Our goal in this section 
is to calculate the function $\Omega_{\text{gw}}(\nu)$ for the 
gravitons produced during the Big Bounce phase.

We consider the creation of the gravitons during the transition from some initial to final states. 
The initial vacuum state $| 0_{\text{in}}\rangle$  is determined by 
$\hat{a}_{\text{k}}| 0_{\text{in}}\rangle = 0$, where  $\hat{a}_{\text{k}}$ is the initial annihilation operator for $\tau_i$. 
The relation between annihilation and creation operators for the initial and final states is given 
by the Bogoliubov transformation      
\begin{eqnarray}
\hat{b}_{{\bf k}} &=& B_{+}(k) \hat{a}_{{\bf k}} + B_{-}(k)^{*}  \hat{a}_{-{\bf k}}^{\dagger} \ , \label{Bog1}  \\
\hat{b}_{{\bf k}}^{\dagger} &=& B_{+}(k)^{*}\hat{a}_{{\bf k}}^{\dagger} + B_{-}(k)    \hat{a}_{-{\bf k}} \label{Bog2} 
\end{eqnarray}
where $|B_{+}|^2-|B_{-}|^2=1$. Because we are working in the Heisenberg description the vacuum state does not 
change during the evolution. It results that
$\hat{b}_{{\bf k}}| 0_{\text{in}}\rangle=B_{-}(k)^{*}  \hat{a}_{-{\bf k}}^{\dagger}| 0_{\text{in}}\rangle $  is 
different from zero when $B_{-}(k)^{*}$ is a nonzero function. This means that in the final state  
graviton field considered is no more in the vacuum state without particles. The number of produced particles in the 
final state is given by  
\begin{equation}
\bar{n}_{{\bf k}} = \frac{1}{2} \langle 0_{\text{in}} |\left[ \hat{b}_{{\bf k}}^{\dagger}\hat{b}_{{\bf k}}+
 \hat{b}_{-{\bf k}}^{\dagger}\hat{b}_{-{\bf k}} \right]| 0_{\text{in}} \rangle =|B_{-}(k)|^2. \label{particles}
\end{equation}
The energy density of gravitons is given by   
\begin{equation}
d\rho_{\text{gw}} = 2 \cdot \hslash \omega \cdot  \frac{4 \pi \omega^2   d\omega}{(2\pi c)^3} \cdot |B_{-}(k)|^2.
\end{equation}
where we used definition (\ref{particles}). The expression for the parameter $\Omega_{\text{gw}}$ defined by (\ref{omegaGW})
takes now the form
\begin{equation}
\Omega_{\text{gw}}(\nu) = 
\left\{ \begin{array}{ccl} 
0 & \mbox{for}  & k>k_0  \\
\Omega_0  \cdot \nu^4 \cdot \sinh^2\left[  \sqrt{1- 
 \left(k/k_0 \right)^2 } (\tau_{\text{i}}-\tau_{\text{f}})k_0 \right]   & \mbox{for}  & k\leq k_0
\end{array} \right. 
\end{equation}
where 
\begin{equation}
k = 2\pi \nu \frac{a_\text{0}}{a_{\text{f}}},
\end{equation}
and
\begin{equation}
\Omega_0 = \frac{\hslash c}{c^4}\frac{16\pi^2}{\rho_*}
 = 3.66 \cdot h^{-2}_0 \cdot 10^{-49} \ [\text{Hz}^{-4}].  
\end{equation}
For the model with the inflationary phase we have $\frac{a_\text{0}}{a_{\text{f}}} \simeq 10^{56}$.
Another value which must be specified is the duration of the bounce. We assume 
that $\tau_{\text{i}}=-20 \ l_{\text{Pl}}$ and the  $\tau_{\text{f}}=20 \  l_{\text{Pl}}$.
  
\begin{figure}[ht!]
\centering
\includegraphics[width=9cm,angle=0]{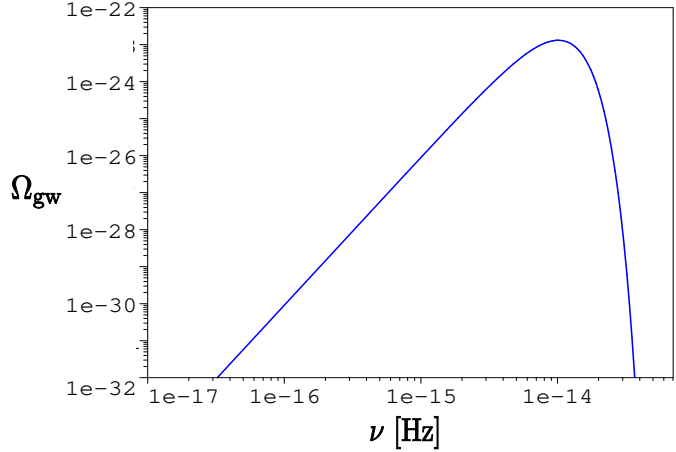}
\caption{Function $\Omega_{\text{gw}}(\nu)$ for the 
$\tau_{\text{i}}=-20 \ l_{\text{Pl}}$ and the  $\tau_{\text{f}}=20 \  l_{\text{Pl}}$.}
\label{Omega}
\end{figure}

In Fig. \ref{Omega} we show the function $\Omega_{\text{gw}}(\nu)$ for the setup considered. 
The spectrum is characterised by the bump with the maximal frequency 
\begin{equation}
\nu_{\text{max}} \simeq 7 \cdot 10^{-14} \ [\text{Hz}].
\end{equation}

The presented result approves our previous consideration. Namely, there is no production 
of the gravitons with frequencies above some $\nu_{\text{max}} $. The only contribution
to the spectrum comes from the bump.   

\section{Tensor modes from the pre-bounce phase}

The bouncing solution (\ref{bouncesol}) in the limit $t\rightarrow \pm \infty$ gives 
\begin{equation}
a(t) \propto |t|^{1/3} \propto |\tau|^{1/2}.
\end{equation}
In this regime the correction $T_Q$ vanishes and the expression for the effective mass simplifies to 
\begin{equation}
m^2_{\text{eff}}=-\frac{a^{''}}{a} = \frac{1}{4} \frac{1}{\tau^2}.
\end{equation}
The equation for the mode functions takes a form 
\begin{equation}
\frac{d^2}{d\tau^2}f(k,\tau) + \left[ k^2+\frac{1}{4}\frac{1}{\tau^2}  \right] f(k,\tau) = 0, 
\end{equation}
The normalised solution of this equation has a form
\begin{equation}
f(k,\tau) =   \frac{\mathcal{N}}{\sqrt{2k}} \sqrt{-\tau k} H^{(1)}_{0}(-\tau k) \label{SlowRollSol1}
\end{equation}
where
\begin{equation}
\mathcal{N} = \sqrt{\frac{\pi}{2}} e^{i\pi/4}.
\end{equation}
With the use of definitions (\ref{PowerSpectDef}) and (\ref{SlowRollSol1}) 
we obtain the power spectrum for the pre-bounce phase
in the form
\begin{equation}
\mathcal{P}_T(k) = \sqrt{\frac{12}{\pi}} |H_0^{(1)}(2\pi)|^2  \left( \frac{k}{k_{\#}} \right)^3 
\ \ \text{for} \ \ k \rightarrow 0,
\end{equation}
where $k_{\#}$ is some constant.  

It is important to note that this part of the spectrum does not depend on the 
quantum gravitational effects. It is in opposite to the predictions from the 
inflationary models, where low energy modes come from a high energy region. 
Here, low energy modes are produced in the low energy pre-Big Bang state.
So, the present largest scale structures have their origin 
in the semi-classical pre-Big Bang phase rather than in the 
deep quantum regime. 

\section{Suppressing low  CMB multipoles with a bounce+inflation scenario}

\begin{figure}[ht!]
\centering
\includegraphics[width=9cm,angle=0]{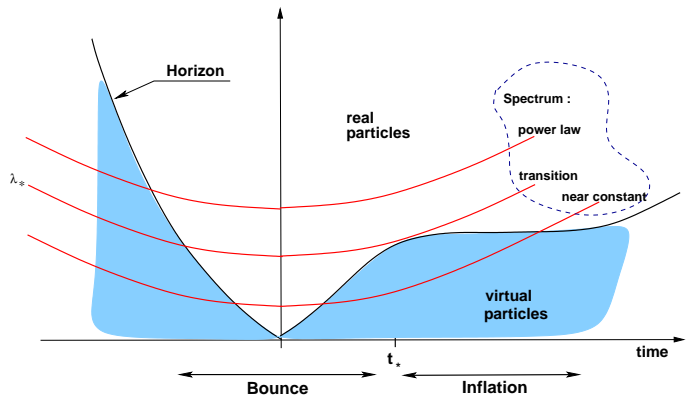}
\caption{Evolution of the horizon in the bounce+inflation model.}
\label{BI}
\end{figure}

In this section we consider scenario with an inflationary phase taking place after   
the bounce. In Fig. \ref{BI} we show the resulting evolution of the horizon.
We see that the horizon is firstly crossed by the modes of the length $\lambda>\lambda_*$. 
These modes lives frozen on the super-horizontal scales and re-enter the horizon
in the post-Big Bang phase. Additionally they are not sensitive for the quantum 
gravitational effects close to the bounce. As it was shown in the previous section 
they leads to the power spectrum in the form 
\begin{equation}
\mathcal{P}_T(k)  \propto k^3 \ \  \text{for}   \ \ k \ll k_*.
\end{equation}
These fluctuations can lead directly to the CMB fluctuations on the large angular scales. 
Modes of the length $\lambda<\lambda_*$ also cross the horizon in the pre-Big Bang phase but 
re-enter shortly after bounce. They decay shortly after this and become a quantum 
fluctuations. Then, during the inflationary phase they cross the horizon again.
The spectrum of these fluctuations will be nearly flat as predicted by the 
inflationary models 
\begin{equation}
\mathcal{P}_T(k) = \mathcal{A} \ \ \text{for} \ \    k \gg k_*.
\end{equation}
The loop holonomy corrections to the inflationary spectrum were studied in Ref. 
\cite{Zhang:2007bi,Artymowski:2008sc}.

\begin{figure}[ht!]
\centering
\includegraphics[width=7cm,angle=0]{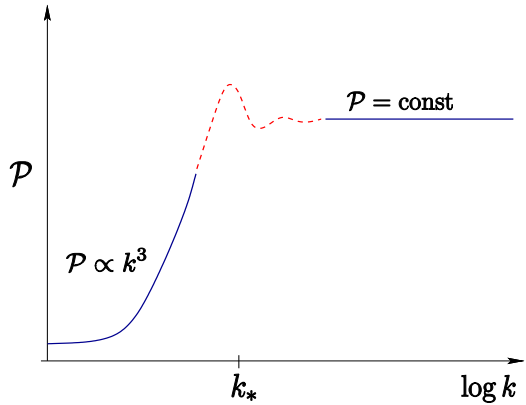}
\caption{The schematic picture of the power spectrum expected for the bounce+inflation model.}
\label{BISpect}
\end{figure}

Finally, we should obtain the spectrum of the tensor perturbations in the form 
shown in Fig. \ref{BISpect}. 

\begin{figure}[ht!]
\centering
\includegraphics[width=7cm,angle=0]{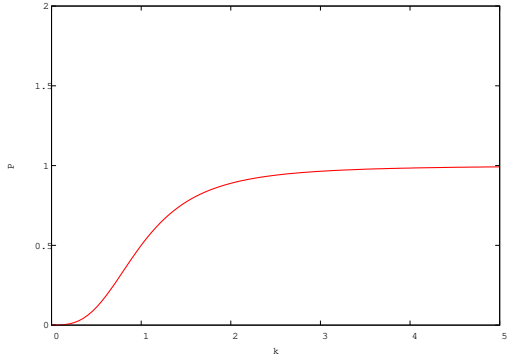}
\caption{Phenomenological spectrum for the bounce+inflation model with $\mathcal{A}=1$ and $k_*=1$.}
\label{PhenSpect}
\end{figure}

To the describe power spectrum  in a continuous way we propose the phenomenological formula
\begin{equation}
\mathcal{P}(k) = \mathcal{A} \left( \frac{k}{k_*}\right)^3 \frac{1}{1+\left( \frac{k}{k_*}\right)^3}, \nonumber
\end{equation}
which interpolates the two regimes considered. We plot this function in Fig. \ref{PhenSpect}.

The spectrum in the same form should be also obtained for the scalar modes of fluctuation.   
In this context similar models were studied in Ref. \cite{Contaldi:2003zv,Piao:2003zm}. They have 
shown that a transitional regime in the spectrum can have form of the oscillations, 
as it was shown in Fig. \ref{BISpect}. However, in the cited papers evolution was
obtained form the specific dynamics of the single inflationary field. Namely, no quantum
gravitational corrections was used to obtain the bounce phase.   

An interesting property of the obtained spectra is damping of the low multipoles in the 
CMB spectra. Such an effect is in fact observed \cite{Shafieloo:2003gf}. This gives 
potentially the possibility to test bouncing cosmologies. If the bounce is present 
we should observe low multipoles suppression. However, this effect can have also 
other origins and on the present stage of observations it is not possible to 
indicate that it is truly the remnant of the Big Bounce.      
 
The important feature of the presented scenario is that none of the primordial perturbations 
produced in the deep quantum gravitational regime gives seeds to the structures formation.
All perturbations come from either the pre-Big Bang semi-classical phase or from the  
post-Big Bang inflationary phase. It is therefore hard to investigate observationally 
the deep quantum regime. However, we can potentially observe the classical pre-Big Bounce 
branch which is the result of the quantum gravitational effects.  

\section{Summary}

In this paper we have considered the gravitational waves creation during the Big Bounce phase inspired
by the Loop Quantum Cosmology. We have studied effects of the holonomy corrections to the equation
for tensor modes. We have shown that they lead to the suppression of the gravitons production. 
However, this effects is not dominant and can be treated perturbatively. In fact, to obtain 
qualitative results  it is justified to neglect this contribution. Based on the above studies we have 
solved the simplified model of the gravitons production during the bounce phase. We have 
derived values the power spectrum $\mathcal{P}_T$ and the parameter $\Omega_{\text{gw}}$. 
The obtained spectrum for the bounce has a form of the bump. It decreases to zero for the 
energies tending to zero and for some high energy scale. 
This is a typical property of the bouncing cosmologies. The similar 
behaviour is also expected for the scalar perturbations. The perturbations 
form the pre-Big Bang phase gives a direct imprint for the large scale structures. 
In particular they can lead to the suppression of the low CMB multipoles. 
However, the bounce phase itself does not lead to observed features of the CMB spectra.       
The post-bounce inflationary phase is required. Such a model 
with the bounce and the following inflationary phase is not yet constructed 
in the Loop Quantum Cosmology. It is therefore a challenge to construct it in a 
consistent way. 

\begin{acknowledgments}
I am grateful to prof. Marek Szyd{\l}owski and dr Adam Krawiec for careful reading the manuscript.
This work has been supported by the Marie Curie Host Fellowships for the
Transfer of Knowledge project COCOS (Contract No. MTKD-CT-2004-517186).
\end{acknowledgments}

\end{document}